\documentclass[aps,pre,twocolumn,final,10pt,groupedaddress,showpacs]{revtex4-1}
\usepackage{amsmath}
\usepackage{amsfonts}
\usepackage{amssymb}
\usepackage{color}
\usepackage{epstopdf}
\usepackage{bm}
\usepackage{natbib}
\usepackage[pdftex]{graphicx}
\usepackage{times}
\usepackage{txfonts}
\usepackage{textcomp}
\usepackage{bbold}
\usepackage[colorlinks, bookmarks=false, citecolor=blue, linkcolor=red, urlcolor=blue]{hyperref}

\begin{document}
\title{Linear Stability and the Braess Paradox in Coupled Oscillators Networks and Electric Power Grids}
\author{Tommaso Coletta}
\affiliation{School of Engineering, University of Applied Sciences of Western Switzerland, CH-1951 Sion, Switzerland}
\author{Philippe Jacquod}
\affiliation{School of Engineering, University of Applied Sciences of Western Switzerland, CH-1951 Sion, Switzerland}
\date{\today}

\begin{abstract}
We investigate the influence that adding a new coupling has on the linear stability of the synchronous state 
in coupled oscillators networks.
Using a simple model we show that, depending on its location, the new coupling can lead to enhanced or reduced stability.
We extend these results to electric power grids where a new line can lead to four different scenarios 
corresponding to enhanced or reduced grid stability as well as increased or decreased power flows.
Our analysis shows that the Braess paradox may occur in any complex coupled system, 
where the synchronous state may be weakened and sometimes 
even destroyed by additional couplings.
\end{abstract}

\pacs{05.45.Xt, 88.80.hh, 88.80.hm}

\maketitle

\section{Introduction.} 
Collective synchrony is an omnipresent phenomenon in systems of coupled 
oscillators~\cite{Kuramoto75,Strogatz01}. It arises when the coupling
between individual oscillators becomes strong enough that it overcomes the 
tendency of oscillators to swing at their natural frequencies.
Simplified models such as the Kuramoto model~\cite{Kuramoto75,Acebron05} allow to describe a plethora of non linear 
phenomena involving collective synchrony in
Josephson junction arrays~\cite{Wiesenfeld}, biological systems~\cite{Winfree,Ermentrout}, crowd dynamics~\cite{Strogatz05}, 
coupled neural networks~\cite{Arbib02}, chemical reactions~\cite{Kuramoto84} and electric power grids~\cite{Dorfler12,Dorfler13,
Motter,Backhaus13} to name but a few. 
Quite naturally, one expects that adding couplings between initially 
uncoupled pairs of oscillators generically favors synchrony. This is however not
always the case, Nishikawa and Motter provided analytical conditions for systems of coupled oscillators to be synchronizable over 
a larger parameter range \cite{Nishikawa2006}. 
References~\cite{Witthaut-NJP,Witthaut-EPJ} found numerically that 
adding a new coupling in an initially synchronous system sometimes destroys synchrony.
This unexpected scenario is the electrical analog 
of the Braess paradox,
first discussed in the context of traffic networks~\cite{Braess1,Braess2}, where
building new roads sometimes increases traffic congestions.
Similar counterintuitive observations were reported for simple mechanical systems and 
uncontrolled electric circuits~\cite{Cohen,Blumsack}.
One purpose of the present manuscript is to present a more systematic
analytical treatment of the Braess paradox in coupled oscillator systems. While our focus is
on electric power grids, our theory also applies to other oscillator networks described by similar models.

The operational state of AC power grids requires 
synchrony of thousands of rotating machines of widely varying sizes,
millions of electric and electronic devices and components, over several voltage levels
intercoupled by frequency-preserving transformers~\cite{Kundur}. 
Nowadays, power grids are maintained in a synchronous state at their rated frequency ($50$ or $60$ Hz)
by active control of power generators.
An inbalance between production and consumption results in a variation of the operating frequency
but not necessarily in the loss of synchrony. The latter may arise if frequency variations exceed safety margins
that require to disconnect parts of the network.
The standard operational protocol is crucially challenged by the current
rise of weakly controllable renewable energy sources.
Maintaining the operational state and guaranteeing  
the safe distribution of power under these changing circumstances requires power grid upgrades,
in particular the addition of new transmission lines. This is however both costly 
and not always well accepted socially.
It is therefore crucial to upgrade grids efficiently, adding as few lines as possible
to ensure a more stable and safer grid operation. Understanding the Braess paradox in electric 
power systems is therefore key to optimize the grid of tomorrow.

Improvement in grid operation after a line addition can be quantified for instance by 
(i) the new power flows on initially strongly loaded lines, this measure of power re-routing is 
 related to line outage distribution factors used in electrical engineering~\cite{Wood},  
(ii) the linear stability as measured by the Lyapunov 
spectrum~\cite{Pecora-PRL,Fink-PRE,Motter,Mallada2011,Wittahaut-EPJS}
of the upgraded grid and (iii) the size of the basin of attraction of the synchronous state
in the associated parameter space~\cite{Menck}.
Further criteria include N-1 feasibility and voltage stability \cite{Kundur}.
In this work we investigate the impact of line addition on grid operation along points (i) and (ii) 
in a purely reactive power grid.
We illustrate analytically on a simple chain network how the perturbative addition of a line, 
which modifies the grid topology by creating a loop, 
affects the power load of the electrical connections and the linear stability.
We classify the impact of the new line into one of four different scenarios - depending on whether linear stability is improved or 
not, and whether strongly loaded power lines are relieved or not.
Out of these four possible scenarios, three are different manifestations of the electrical Braess paradox, where (I) already
strongly loaded lines become even more strongly loaded, (II) network stability is reduced or
(III) both. 
We furthermore show how these three scenarios for the Braess paradox also occur in a complex network having 
the topology of the British electric transmission grid.
Our analytical calculation contributes to the understanding of the Braess paradox in electrical systems.
We conjecture that the paradox is generic and may occur 
in any system of coupled oscillators with reduced connectivity.

\section{The chain model.}
\begin{figure}[htbp]
 \centering
 \includegraphics[width=\columnwidth]{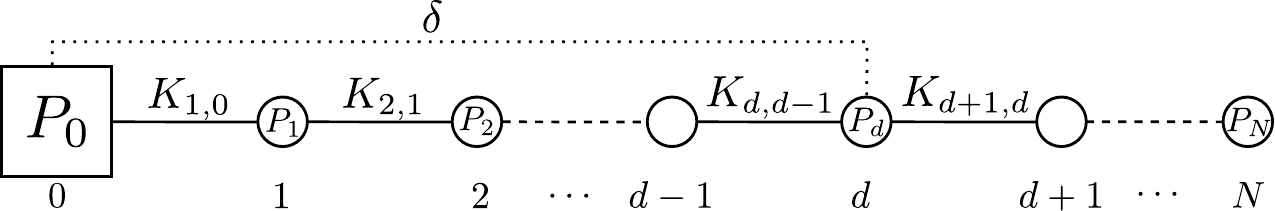}
 \caption{The chain model. 
 A single generator (square) injects a power $P_0>0$ which is consumed by $N$ loads (circles)
 each consuming a power of $P_i<0$ in arbitrary units. 
 The lines have capacity $K_{i+1,i}\geq P_0-\sum_{l=1}^i |P_l|$,
 except the newly added line (dotted) which has capacity $\delta$.}
 \label{fig:Network}
\end{figure}

We consider an AC electric power system in the form of a chain connecting $N+1$ nodes [see Fig.~\ref{fig:Network}].
A unique generator (labeled $i=0$) is located at one end of the branch while the remaining $N$ nodes
(labeled $i=1,\ldots,N$) are all loads. 
A necessary condition for the system to be in steady state is that the total injected power at the generator
is equal to the total power consumed by the loads. 
For a power injection $P_0>0$ at the generator and $P_i<0$ at the loads, in arbitrary units, this amounts to 
$\sum_{i=0}^NP_i=0$.
As is the case for high voltage transmission grids, the line admittance is dominated by the susceptance (its imaginary part). 
Accordingly we neglect ohmic effects and assume that each node has a constant internal voltage magnitude $|V_i|$.
Under these approximations the active power flow equations read~\cite{Kundur,Arenas2008,Filatrella2008,Chopra09,Witthaut-NJP,Delabays2015}
\begin{equation}\label{eq:PowerFlow}
 0=P_i+\sum_{j \sim i}K_{i,j}\sin\left(\theta_{j}-\theta_i\right)  \quad i\in\left\{0,1,\ldots,N\right\} \,,
\end{equation}
where $j \sim i$ indicates that the sum over $j$ spans the nodes connected to the $i^\textrm{th}$ node,
$K_{i,j}=K_{j,i}=B_{i,j}|V_i||V_j|$ denotes the maximum power capacity of line $\langle i,j\rangle$
having susceptance $B_{i,j}$ and $\theta_i$ is the voltage angle (with respect to the 
current) at node $i$.
Since $P_i^\star=P_0-\sum_{l=1}^i |P_l|$ units of power are transmitted from node $i$ to $i+1$,
the line capacities must satisfy $K_{i+1,i}\geq P_i^\star$ for Eqs.~(\ref{eq:PowerFlow}) to have a solution.
Solving Eqs.~(\ref{eq:PowerFlow}) for the angles yields
\begin{equation}\label{eq:angles}
\theta_{i+1,i}\equiv\theta_{i+1}-\theta_{i}=-\arcsin\left[P_i^\star/K_{i+1,i}\right] \quad i\in\{0,1,\ldots,N-1\}\,, 
\end{equation}
so that $\theta_{i+1,i}\in[-\pi/2,0]$.

We next add to this topology a line of capacity $\delta$
between the generator and the $d^{\rm th}$ load, $d\in\{1,\ldots,N\}$.
When $\delta \ll K_{i+1,i}$ for all $i$, the perturbed solution 
$\{\tilde{\theta}_i\}$ remains close to the unperturbed one, i.e. 
$\tilde{\theta}_{i+1,i}\approx\theta_{i+1,i}+\epsilon_{i+1,i}$ with $|\epsilon_{i+1,i}|\ll1$. 
Solving for the $\epsilon_{i+1,i}$'s, one obtains explicitly the $1^\textrm{st}$ order correction to the 
unperturbed power flow solution as
\begin{equation}\label{eq:Full angle solution}
 \epsilon_{i+1,i}=\left\{
 \begin{array}{ll}
\displaystyle -\frac{\delta}{K_{i+1,i}}\frac{\sin\theta_{d,0}}{\cos{\theta_{i+1,i}}}  \quad & 0\leq i\leq d-1\,, \\[3mm]
\displaystyle  0 \quad & i\geq d\,.
  \end{array}\right. \quad
\end{equation}
Clearly, $\epsilon_{i+1,i}=0$ for $i\geq d$ since the power flowing through the lines connecting 
nodes $i$ and $i+1$ for $i\geq d$ is left unchanged.
Since the above result is valid to $1^\textrm{st}$ order in $\delta/K$, 
all angle differences entering Eq.~(\ref{eq:Full angle solution}) are differences of the 
unperturbed angles $\{\theta_i\}$. In particular, the difference between the voltage angles of the 
nodes which are connected by the new line is
$\theta_{d,0}=\sum_{i=0}^{d-1}\left(\theta_{i+1}-\theta_i\right)=-\sum_{i=0}^{d-1}\arcsin\left[P_i^\star/K_{i+1,i}\right] \, .$
Below we investigate how the power flowing through the lines changes when adding the new line.

\section{Impact of line addition on power flows.} 
To leading order in $\delta/K$,  the power flowing through the additional 
line is $P_{d,0}=\delta\sin\theta_{d,0}$ 
\footnote{In our convention $P_{i,j}=K_{i,j}\sin{(\theta_{i}-\theta_j)}$, 
$P_{i,j}>0$ (respectively $P_{i,j}<0$) means that 
a power of $|P_{i,j}|$ is flowing from node $i$ to node $j$ (respectively from node $j$ to node $i$)}.
The sign of $P_{d,0}$, and thus the direction of the power flow, changes as a function
of $d$. 
The power flowing through the $\langle 0,1\rangle$ line between the generator and the first node goes from 
$P_{1,0}= K_{0,1}\sin{\theta_{1,0}}=-P_0$ to
$\tilde{P}_{1,0}= K_{1,0}\sin{\tilde{\theta}_{1,0}}\approx -P_0-\delta\sin{\theta_{d,0}}$ once 
the new line is added.
As long as $\sin\theta_{d,0}\leq0$, the new line lowers 
the load on all the lines $\langle i,i+1 \rangle$ for $i=0, \dots,d-1$.
However, when $\sin\theta_{d,0}\geq0$, we face the counterintuitive situation where the new
line transmits power back from node $d$ to the generator, thereby increasing the load on all 
the lines between the generator and node $d$ in the original network. 
This is an electric manifestation of the Braess paradox~\cite{Braess1,Braess2} and its
occurrence is due to the nonlinear nature of the power flow Eqs.~(\ref{eq:PowerFlow}). In 
the case of our simple model, which of these two scenarios takes place depends only 
on the value of $\theta_{d,0}$.

\section{Linear stability.}
The solutions of the power flow Eqs.~(\ref{eq:PowerFlow}) describe the operating stationary
state of the power grid at a given time. Upon changing conditions, such as variations of the power
injected and consumed, the angles' dynamics in this transient stability problem is governed by
the {\it swing equations}~\cite{Kundur,Backhaus13}
\begin{equation}\label{eq:Swing Eqs}
 I_i \ddot{\theta}_i + D_i\dot{\theta}_i =P_i+\sum_{j \sim i}K_{i,j}\sin\left(\theta_{j}-\theta_i\right)\,,
\end{equation}
which describe the power balance at nodes with rotating machines as generators or loads.
Without inertia $I_i\equiv0$, Eqs.~(\ref{eq:Swing Eqs}) reduce to a Kuramoto-like 
model~\cite{Kuramoto75,Strogatz00,Acebron05}, with reduced connectivity.
Linear stability in the Kuramoto and similar models with reduced connectivity has been investigated in 
Refs.~\cite{Jadbabaie04,Mirollo05,Chopra09,Dorfler2012}, which derived bounds on 
the exponential rate of return to the stationary state.

For $I_i\neq0$, linearizing Eq.~(\ref{eq:Swing Eqs}) around a stationary solution ${\bm\Theta}(t)={\bm\theta}+\bm{\delta\theta}(t)$ yields the eigenvalue equation
\begin{equation}\label{eq:Eigenvalue Eq}
 M\bm{\delta\theta}=\Lambda\left[\Lambda \textrm{diag}({\bf I})+\textrm{diag}({\bf D})\right]\bm{\delta\theta}\,,
\end{equation}
where $\Lambda\rm{'s}\in\mathbb{C}$ are the Lyapunov exponents of the dynamics governed by 
Eq.~(\ref{eq:Swing Eqs}), $\textrm{diag}({\bf I})$
and $\textrm{diag}({\bf D})$ are diagonal matrices,  $\textrm{diag}({\bf I})_{ii} = I_i$ 
and $\textrm{diag}({\bf D})_{ii} = D_i$. Finally, $M$ is the stability matrix defined by
$M_{ij}=K_{i,j}\cos{\theta_{j,i}}$ if $i$ and $j$ are connected,
$M_{ii}=-\sum_{l\sim i} M_{i,l}$ and zero otherwise~\cite{Motter,Pecora-PRL}.
The stationary solution is linearly stable if the largest nonzero Lyapunov exponent is negative and
unstable otherwise. We next show that the system is stable if $M$ is negative semidefinite and that the
loss of stability occurs when the largest nonzero eigenvalue of $M$ becomes positive. 
This justifies our use of the spectrum of $M$ as measure of stability, keeping in mind that time scales, e.g. for
restoring synchrony may depend on $I_i$ and $D_i$.

In the case of homogeneous inertia $I_i\equiv I$ and damping coefficients $D_i\equiv D$,
Eq.~(\ref{eq:Eigenvalue Eq}) is diagonalized by the eigenvectors of $M$ and the Lyapunov exponents
are simply given by
\begin{equation}\label{eq:Stability with inertia}
 \Lambda_a^\pm=-\frac{\beta}{2}\pm\frac{1}{2}\sqrt{\beta^2+4\lambda_aI^{-1}}\,,
\end{equation}
where $\beta=D/I$ and $\lambda_a$ is one of the eigenvalues of the stability matrix $M$.
In the inhomogeneous case, projecting Eq.~(\ref{eq:Eigenvalue Eq}) onto $\bm{\delta\theta}$
gives,
\begin{equation}\label{eq:Eigenvalue Eq projected}
 0=\Lambda^2 a+\Lambda b-c\,,
\end{equation}
where we introduced the shorthand notation for the overlaps
$a=\bm{\delta\theta}\textrm{diag}({\bf I})\bm{\delta\theta}$, $b=\bm{\delta\theta}\textrm{diag}({\bf D})\bm{\delta\theta}$ 
and $c=\bm{\delta\theta}M\bm{\delta\theta}$, with coefficients $a,b\geq0$ since inertia and damping coefficients are positive quantities (i.e. $D_i,I_i\geq0~\forall i$).
The Lyapunov exponents then take the form
\begin{equation}\label{eq:Lyapunov inhomogeneous}
 \Lambda^\pm=\frac{1}{2a}\left(-b\pm\sqrt{b^2+4ac}\right)\,.
\end{equation}
In both Eqs.~(\ref{eq:Stability with inertia}) and (\ref{eq:Lyapunov inhomogeneous}), $\textrm{Re}[\Lambda^-]$ is always negative and 
stability depends on the sign of $\textrm{Re}[\Lambda^+]$.

In the homogenous case,
Eq.~(\ref{eq:Stability with inertia}) makes it clear  
that linear stability is determined uniquely by the spectrum of $M$: $\textrm{Re}[\Lambda_a^+]$
and $\lambda_a$ become positive simultaneously.
Since $M$ is real and symmetric, all $\lambda_a$'s are real. Thus, the necessary condition
for the system to be stable is that all $\lambda_a$ are negative. 
Furthermore, $\textrm{Re}[\Lambda_a^+]$ is negative and decreasing as $\lambda_a$ decreases in the interval $0>\lambda_a>-I\beta^2/4$, while 
$\textrm{Re}[\Lambda_a^+]$ saturates at $-\beta/2$ when $\lambda_a$ is decreased further. Thus,
not too far from loss of stability, the spectrum of $M$ is as good a measure of increase/decrease
of stability as the true Lyapunov spectrum.

We extend the approach of Ref.~\cite{Wittahaut-EPJS}, which deals with inhomogeneous damping but identical inertia, to the case of inhomogenous inertia and damping.
When $M$ is negative semidefinite, the coefficient $c$ in Eq.~(\ref{eq:Lyapunov inhomogeneous}) is negative. 
Thus, given that $a\geq0$, $\textrm{Re}[\Lambda^+]$ is negative and the solution is linearly stable. 
Furthermore, the cancellation of the Lyapunov exponent occurs only when $c$ vanishes. 
For this to take place, $\bm{\delta\theta}$ must be proportional to 
one of the eigenvectors of $M$ associated to a zero eigenvalue.
This shows that 
a stationary solution of the dynamic system (\ref{eq:Swing Eqs}) is linearly stable as long as 
$M$ is negative semidefinite and the loss of stability occurs when the largest nonzero eigenvalue of $M$ vanishes.

%

We therefore take from now on the spectrum
of $M$ as a measure for increased ($\lambda_a$ decreases)
or decreased stability ($\lambda_a$ increases). 
Because inertia does not influence the stationary state, taking this latter criterion allows to make more
general statements regarding stability, however one needs to keep in mind that $I_i$
and $D_i$ may in principle affect stability in a nontrivial way. We defer investigations of this issue
to future work.

Having discussed the role of the spectrum of the stability matrix on the Lyapunov exponents we
next investigate how stability is affected as a line is added to an initially stable network.

\section{Linear stability for the chain model}
In the case of the chain model prior to line addition, $M$ is a $(N+1)\times(N+1)$ tridiagonal, 
symmetric matrix.
 \begin{equation}\label{eq:Stability matrix unperturbed case}
 M=-\left(
  \begin{array}{cccccc}
   C_{1,0} &  -C_{1,0}     &   0     & \ldots & 0 \\
  -C_{1,0} &C_{1,0}+C_{2,1} & -C_{2,1} &   0    & \vdots \\
     0    & \ddots       & \ddots  & \ddots & 0      \\
  \vdots  &    0         & \ddots  & \ddots & -C_{N,N-1} \\
     0    & \ldots       &    0    &-C_{N,N-1} & C_{N,N-1}
  \end{array}
  \right)\,,
 \end{equation}
where $C_{j,i}\equiv K_{j,i}\cos\theta_{j,i}$. 
Since $\theta_{i+1,i}\in[-\pi/2,0]$, 
$M$ is diagonally dominant~\cite{Horn_MatrixAnalysis} with only negative diagonal elements and positive subdiagonal elements. It
thus belongs to the family of Jacobi matrices~\cite{Jacobi_matrices}, in particular $M$ has
distinct eigenvalues.
By Gershgorin circle theorem
\footnote{Gershgorin circle theorem ensures that every eigenvalue of a matrix $M$ is contained in at least one of the
discs centered at $M_{i,i}$ and of radius $\sum_{j\neq i}|M_{i,j}|$ (see Ref.~\cite{Horn_MatrixAnalysis}).}, it is negative 
semi-definite.
Furthermore, $\mathbf{u}^{(1)}=(1,\ldots,1)$ is the eigenvector associated to the eigenvalue which 
vanishes by rotational invariance.
We order the eigenvalues of $M$ as $\lambda_1=0>\lambda_2>\ldots>\lambda_{N+1}$.
The semi-negativity of the stability matrix indicates that the power flow solution of the original network topology 
is stable against small perturbations. 
Since the largest eigenvalue $\lambda_1$ vanishes,
stability is determined by $\lambda_2$.
Next, we therefore 
calculate the leading order correction to $\lambda_2$ resulting from the line addition. 

The stability matrix $\tilde{M}$ after the new line has been added, has a very similar structure to $M$ except that,
first, the angles entering in $\tilde{M}$ are the $\tilde{\theta}_i$'s and  
second, the new line modifies the following matrix elements: 
$\tilde{M}_{1,1}=-\tilde{C}_{1,0}-\tilde{C}_{d,0}$, $\tilde{M}_{d+1,1}=\tilde{M}_{1,d+1}=\tilde{C}_{d,0}$ 
and $\tilde{M}_{d+1,d+1}=-\tilde{C}_{d-1,d}-\tilde{C}_{d+1,d}-\tilde{C}_{d,0}$, where $\tilde{C}_{j,i} \equiv K_{ij}\cos\tilde{\theta}_{j,i}$
and $K_{d,0}=K_{0,d}=\delta$.
Using Eq.~(\ref{eq:Full angle solution}), 
we express $\tilde{M}$ as $\tilde{M}=M+\Delta M+\mathcal{O}[(\delta/K)^2]$,
where $\Delta M$ is the leading order correction to the stability matrix,
\begin{equation}\label{eq:deltaM1}
 \Delta M=\delta \sin{\theta_{d,0}}
 \left(
 \begin{array}{cc}
  \Delta \mathbb{M}_{(d+1)\times (d+1)} & 0_{(d+1)\times (N-d)}\\
   0_{(N-d)\times (d+1)}            & 0_{(N-d)\times (N-d)}
 \end{array}
 \right)\,,
\end{equation}
with $\Delta \mathbb{M}$ defined as 
\begin{equation} \label{eq:deltaM2}
  \left(
 \begin{array}{cccccc}
-CT_{d,0}-T_{1,0} &    T_{1,0}     &   0    &  \ldots  & CT_{d,0} \\
      T_{1,0}      & -T_{1,0}-T_{2,1}& T_{2,1} &    0     & \vdots \\
       0          &   \ddots      & \ddots &  \ddots  & 0      \\
     \vdots       &      0        & \ddots &  \ddots  & T_{d,d-1}   \\
    CT_{d,0}      &   \ldots      &   0    & T_{d,d-1}&-CT_{d,0}-T_{d,d-1}
 \end{array}
 \right)\,,
\end{equation}
where we introduced the notations $T_{i+1,i}\equiv \tan{\theta_{i+1,i}}$ 
and $CT_{d,0}\equiv \cot\theta_{d,0}$.

Let $\mathbf{u}^{(2)}\in\mathbb{R}^{N+1}$ be defined by $M\mathbf{u}^{(2)}=\lambda_2\mathbf{u}^{(2)}$. 
Then, the leading order correction to $\lambda_2$ is given by 
$\Delta\lambda_2=\mathbf{u}^{(2)\top}\Delta M \mathbf{u}^{(2)}$.
If the sign of $\Delta\lambda_2$ is negative (positive), then, to $1^\textrm{st}$ order in 
$\delta$, the stability of the power flow solution is enhanced (reduced).
Below we discuss how $\textrm{sgn}(\Delta\lambda_2)$ changes as a function of the position $d$ of the additional 
connection. To achieve this, we distinguish the two cases $\tan\theta_{d,0}\geq0$ and $\tan\theta_{d,0}\leq0$.

When $\tan\theta_{d,0}\leq0$, the matrix $\Delta\mathbb{M}$ is diagonal dominant \cite{Horn_MatrixAnalysis}
(since $\theta_{i+1,i}\in[-\pi/2,0]$ we have $\tan\theta_{i+1,i}\leq0$)
with a strictly positive diagonal. Hence $\Delta M$ (\ref{eq:deltaM1}) is either semi-positive or semi-negative definite
depending exclusively on the sign of $\sin\theta_{d,0}$.
In both cases the sign of $\Delta\lambda_2=\mathbf{u}^{(2)^\top}\Delta M \mathbf{u}^{(2)}$ is well defined regardless of $\mathbf{u}^{(2)}$,
and we have
\begin{equation}
\left\{
\begin{array}{l}\label{dlambda1}
 \Delta\lambda_2\geq0 \quad \textrm {for} \quad \theta_{d,0}\in[\pi/2,\pi] \Rightarrow \textrm{reduced stability},\\[2mm]
 \Delta\lambda_2\leq0 \quad \textrm {for} \quad \theta_{d,0}\in[-\pi/2,0] \Rightarrow \textrm{enhanced stability}.
 \end{array}
 \right. \quad
\end{equation}

When $\tan{\theta_{d,0}}\geq0$, $\Delta\mathbb{M}$ is no longer diagonal dominant and it is not 
possible to determine the sign of $\Delta\lambda_2$ as directly as before. 
Instead we use $\mathbf{u}^{(2)}=(u_0^{(2)},\ldots,u_{N}^{(2)})$ to compute $\mathbf{u}^{(2)^\top}\Delta M \mathbf{u}^{(2)}$ explicitly,
\begin{equation}\label{eq:Detail delta lambda_2}
\begin{array}{ll}
 \Delta\lambda_2=&\displaystyle -\delta\cos\theta_{d,0}\left[\left(u_0^{(2)}-u_d^{(2)}\right)^2\right. \\[3mm] 
                 &\displaystyle \left.+\tan\theta_{d,0}\sum_{i=0}^{d-1} \left(u_i^{(2)}-u_{i+1}^{(2)}\right)^2\tan\theta_{i+1,i}\right]\,.
\end{array}
\end{equation}
The one dimensional nature of the model allows to express the difference $u_0^{(2)}-u_d^{(2)}$
as the telescopic sum $\sum_{i=0}^{d-1}(u_i^{(2)}-u_{i+1}^{(2)})=u_0^{(2)}-u_d^{(2)}$. 
Using this identity we rewrite the term $(u_0^{(2)}-u_d^{(2)})^2$ entering Eq.~(\ref{eq:Detail delta lambda_2})
as
\begin{equation}\label{eq:Square of sum}
\begin{array}{ll}
\displaystyle\left[\sum_{i=0}^{d-1}(u_i^{(2)}-u_{i+1}^{(2)})\right]^2=&\displaystyle\sum_{i=0}^{d-1}\left(u_i^{(2)}-u_{i+1}^{(2)}\right)^2 \\
                                                                      &\displaystyle+2\sum_{i=0}^{d-1}\sum_{j>i}(u_i^{(2)}-u_{i+1}^{(2)})(u_j^{(2)}-u_{j+1}^{(2)}) \,.
\end{array}
\end{equation}
Substituting Eq.~(\ref{eq:Square of sum}) in Eq.~(\ref{eq:Detail delta lambda_2}) finally yields
\begin{equation}\label{eq:DeltaLambda2 explicit}
\begin{array}{lll}
\Delta \lambda_2&=&\displaystyle-\delta\cos\theta_{d,0}\left[2\sum_{i=0}^{d-1}\sum_{j>i}(u_{i}^{(2)}-u_{i+1}^{(2)})(u_{j}^{(2)}-u_{j+1}^{(2)})\right.\\[2mm]
& &\displaystyle\left.+\tan\theta_{d,0}\sum_{i=0}^{d-1}(u_{i}^{(2)}-u_{i+1}^{(2)})^2\left(\tan\theta_{i+1,i}+\cot\theta_{d,0}\right)\right]\,.
\end{array}
\end{equation}
Because $M$ is a Jacobi matrix, it can be shown (See Appendix \ref{sec:Jacobi matrices}) that the components of its eigenvector 
$\mathbf{u}^{(2)}$ are monotonously ordered.
Thus $(u_{i}^{(2)}-u_{i+1}^{(2)})(u_{j}^{(2)}-u_{j+1}^{(2)})\geq0$ and the first term in 
Eq.~(\ref{eq:DeltaLambda2 explicit}) is positive.
Furthermore, for $\tan\theta_{d,0}\geq0$, it is possible to establish a sufficient 
condition on $\theta_{d,0}$ according to which the sign of $\Delta\lambda_2$ is known.
Since all $\theta_{i+1,i}$ belong to $[-\pi/2,0]$ and given the monotonicity of the tangent function over this interval,
if $(\tan\theta^{\rm{min}}+\cot\theta_{d,0})\geq0$ where $\theta^{\rm min}={\rm min}_{i\in\{0,\ldots,d-1\}}{\theta_{i+1,i}}$
then we also have $\left(\tan\theta_{i+1,i}+\cot\theta_{d,0}\right)\geq0$ 
for $i\in\{0,1,\ldots,d-1\}$.
When this is the case we conclude that $\Delta\lambda_2\propto -\cos\theta_{d,0}$.
Thus, when $\tan\theta_{d,0}\geq0$, the sign of $\Delta\lambda_2$ is directly given by $-\textrm{sgn}(\cos\theta_{d,0})$ 
if $\tan\theta_{d,0}\leq-\cot\theta^{\rm min}$.
Inspecting Eq.~(\ref{eq:angles}) one sees that $\theta^{\rm min}$ is realized
on the most loaded line prior to the network upgrade, that is on the line having the largest ratio
of transmitted power over available capacity.
Let $\langle q+1,q\rangle$ denote the line minimizing ${\rm min}_{i\in\{0,\ldots,d-1\}}{\theta_{i+1,i}}$
(i.e. the line where $P^\star_i/K_{i+1,i}$ is maximal), then
the condition $\tan\theta_{d,0}\leq-\cot\theta^{\rm min}$ can be rewritten as
\begin{equation}
 \tan\theta_{d,0}\leq\sqrt{K_{q+1,q}^2-{P_q^\star}^2}/P_q^\star\,.
\end{equation}
This defines a critical angle $\alpha\equiv\arctan \Big[\sqrt{K_{q+1,q}^2-{P_q^\star}^2}/{P_q^\star}\Big]\in[0,\pi/2]$, 
such that

\begin{equation}
\left\{
 \begin{array}{l}
  \Delta\lambda_2\leq0 \quad \textrm {for} \quad \theta_{d,0}\in[0,\alpha] \Rightarrow \textrm{enhanced stability},\\[2mm]
  \Delta\lambda_2\geq0 \quad \textrm {for} \quad \theta_{d,0}\in[-\pi,-\pi+\alpha] \Rightarrow \textrm{reduced stability}.
 \end{array}
\right.
\end{equation}
The size of the region $[\alpha,\pi/2]\bigcup[-\pi+\alpha,-\pi/2]$, where the evolution of the stability remains 
undetermined, vanishes as $P_q^\star/K_{q+1,q}$ when $P_q^\star/K_{q+1,q}\rightarrow0$ since in this limit $\alpha\approx\pi/2-P_q^\star/K_{q+1,q}$.

\begin{figure}[htbp]
 \centering
  \includegraphics[width=0.9\columnwidth]{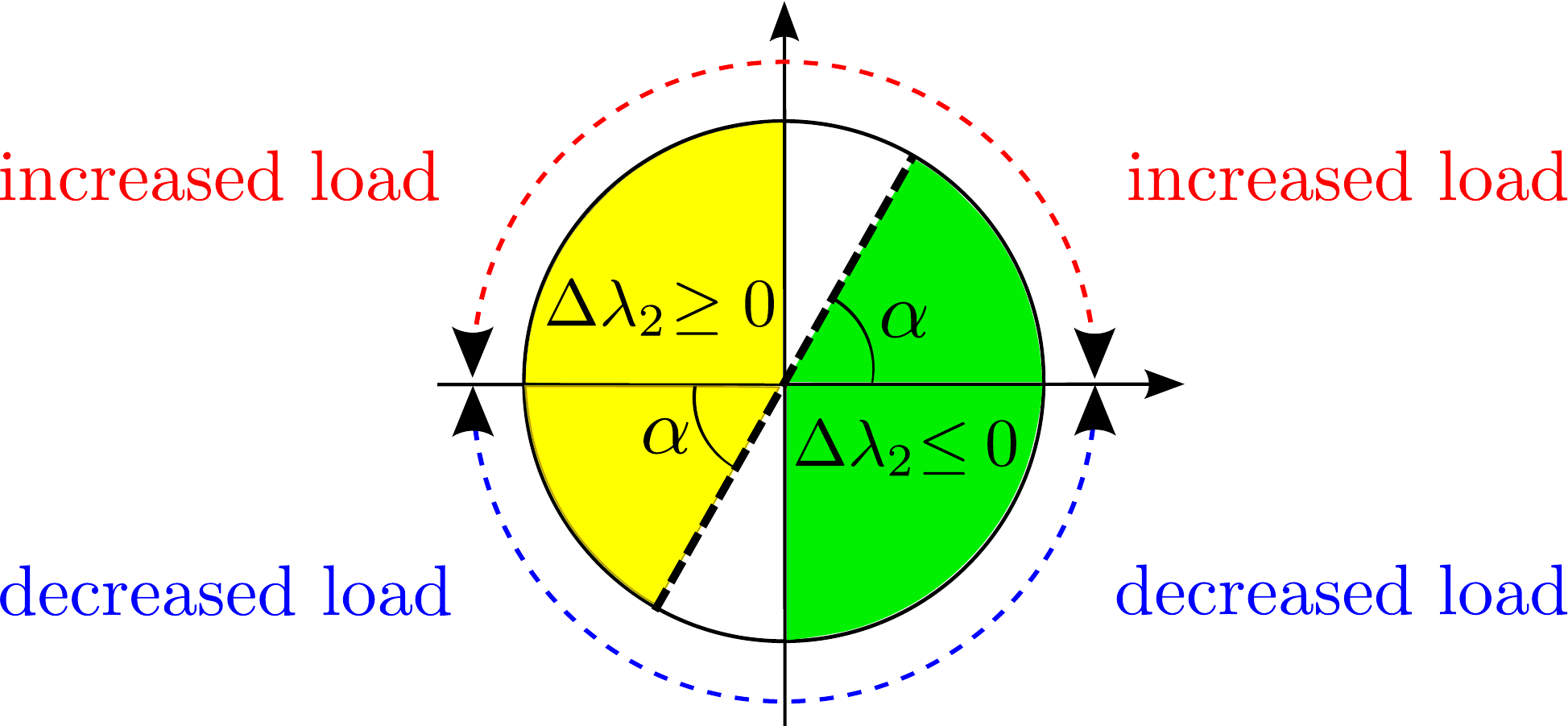}
 \caption{(Color online) Impact of perturbative line addition on the linear stability of the power flow solution 
 (green region, enhanced stability; yellow region, reduced stability) 
 and on the load of the transmission lines (top quadrants, increased load; bottom quadrants,
 decreased loads) as a function of the value of $\theta_{d,0}$. 
}
 \label{fig:Summary}
\end{figure}

These results are summarized in Fig.~\ref{fig:Summary}.
When the angle difference between the newly connected nodes
satisfies $\theta_{d,0} \in [-\pi/2,0]$,
the additional line reduces the load on the most loaded line in the loop (i.e. line $\langle q+1, q\rangle$) and the stability of the 
power flow solution is enhanced. This is what one generally expects of line addition. 
Line addition can however worsen the operating conditions of the network and our
theory highlights three different Braess scenarios how this may happen -- 
by either increasing the 
load, by reducing $|\lambda_2|$, or both.
The worst case scenario occurs when $\theta_{d,0}\in[\pi/2,\pi]$. Then, the additional line 
increases the power load 
on the most loaded line and the stability of the new solution is decreased. 
Paradoxical situations occur 
when $\theta_{d,0}\in[0,\alpha]$ (respectively $\theta_{d,0}\in[-\pi,-\pi+\alpha]$) as the load on the most loaded line
increases (decreases) while the linear stability is enhanced (decreased).
These three outcomes are three different manifestations of Braess's paradox 
in electric power transmission.
We note that the chain model results also apply to the case of line addition in radial (tree like)
networks as long as the new line connects two nodes on the same branch.

\section{Extension to complex networks.}
\begin{figure}[htbp]
 \centering
 \includegraphics[width=\columnwidth]{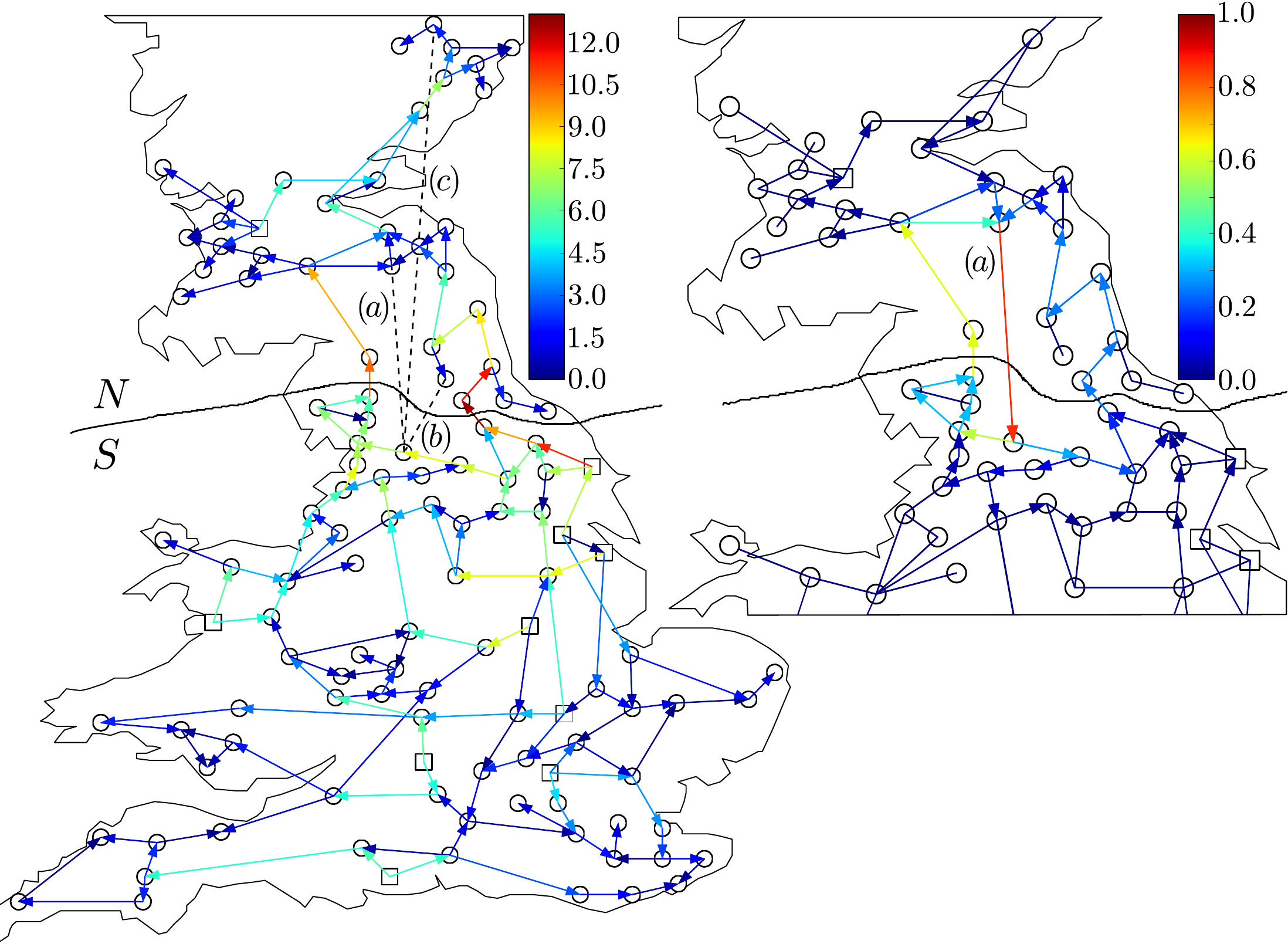}
 \caption{(Color online) Left: UK transmission grid, $10$ generators with power $P=11$
 (squares), $110$ loads with $P=-1$ (circles) and uniform line capacity $K\equiv13$. Power flows are represented by arrows
 and their magnitude is color coded. 
 The dashed lines $(a)-(c)$ represent three different line additions considered and the solid line denotes the network partition 
 into northern and southern zones.
 Right: Plot of the difference in power flows between the solutions after and before the addition of line $(a)$
 of capacity $\delta=1.5$. Arrow heads are drawn only for power flow differences larger than $0.01$.}
 \label{fig:UK}
\end{figure}

To show how the mechanisms described above can lead to the loss of synchrony in more complex networks, we consider the 
electric power transmission grid discussed in Refs.~\cite{Witthaut-NJP,Witthaut-EPJ}. It has the same topology 
as the UK transmission network, and we take the same distribution of loads, 
generators and line capacities as in Ref.~\cite{Witthaut-NJP}.
The general structure of the grid 
is that of a northern, importing zone connected to a southern, exporting
zone via only two lines which are almost at full capacity [see Fig.~\ref{fig:UK}~(left panel)].
It is obviously desirable to relieve these lines by adding another south-north transmission line.

The situation is in a way similar to 
our simple model, where the south plays the role of the generator
and the north that of the loads. This is however only an analogy since 
the elongated UK grid is a meshed network
and not a 1D model as considered above.
In the case of a generic network, Eq.~(\ref{eq:DeltaLambda2 explicit}) becomes
\begin{equation}\label{eq:DeltaLambda2 generic}
\Delta \lambda_2=\displaystyle\delta\sin\theta_{\alpha,\beta}
                 \left[\sum_{\langle i,j\rangle}{f_{ij}\left(u_i^{(2)}-u_j^{(2)}\right)^2}-\left(u_\alpha^{(2)}-u_\beta^{(2)}\right)^2\cot\theta_{\alpha,\beta}\right]\,,
\end{equation}
with
$ f_{i,j}=K_{i,j}\sin\theta_{i,j}\sum_{l\geq2}\left(u^{(l)}_i-u^{(l)}_j\right)\left(u^{(l)}_\alpha-u^{(l)}_\beta\right)\lambda_l^{-1}$ and where
$\alpha$ and $\beta$ are the nodes connected by the new line, $\langle i,j\rangle$ indicates the sum over all pairs of connected 
neighbors in the original network and $\mathbf{u}^{(l)}$ is the $l^\textrm{th}$ eigenvector of the stability matrix 
(see Appendix \ref{sec:Perturbation generic graph}).
After the upgrade, the power flowing through the line connecting nodes $i$ and $j$ becomes 
$P_{i,j}=K_{i,j}\sin\theta_{i,j}+\delta\sin\theta_{\alpha,\beta}~f_{i,j}\cot\theta_{i,j}$.
Following our work, 
a similar expression for the change in power flows resulting from the variation of the capacity of one line
was used in Ref.~\cite{Witthaut2015} to investigate the effect of line failures.

In what follows we present examples of additions of new lines between the north and the south. 
Each illustrates the realization of one of the electric Braess paradoxes
discussed above.
We first add the dashed line $(a)$ [Fig.~\ref{fig:UK}], between two 
nodes having the angle difference $\theta_\textrm{North}-\theta_\textrm{South}\equiv\theta_{N,S}\approx0.9\pi\in [\pi/2,\pi]$. 
For this choice, Fig.~\ref{fig:Summary} predicts 
counterintuitively, that power will flow from the north to the south through the new connection.
This is numerically verified in Fig.~\ref{fig:UK}~(right panel), which shows that
\begin{figure}[htbp]
 \centering
 \includegraphics[width=\columnwidth]{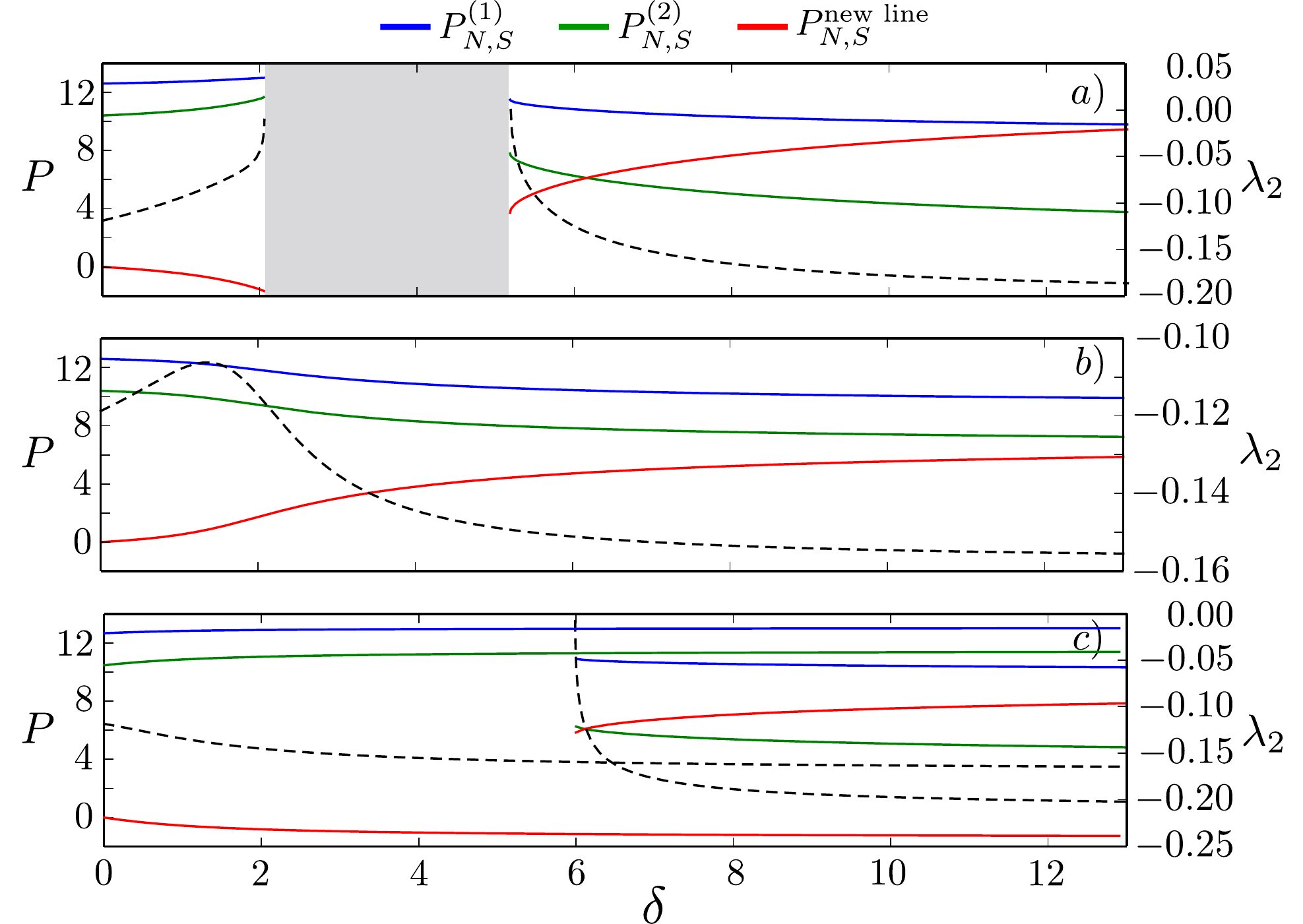}
 \caption{(Color online) Lyapunov exponent (dashed) and power flowing through the lines connecting the north and south areas
 as a function of the capacity of the additional line $\delta$. 
 Each of the panels refers to one of the line additions represented in Fig.~\ref{fig:UK} 
 (labels correspond to the labels of the additional lines in Fig.~\ref{fig:UK})
 and illustrates one of the three Braess scenarios identified in this work.
 Interestingly, panel c) shows the coexistence of two different stable solutions for $\delta\geq5.98$.
} 
 \label{fig:Powers_UK}
\end{figure}
adding the new line increases even further the load on the two original lines 
connecting the two zones - the power flow in the new line goes in
the wrong direction. The effect is quantified 
in Fig.~\ref{fig:Powers_UK}~a) as a function of the capacity of the new line. 
Both the loads on the original connection lines and the 
Lyapunov exponent $\lambda_2$ increase as a function of $\delta$. 
Going beyond the validity of our perturbative approach, synchrony is lost 
in the interval $\delta\in[2.1,5.2]$ (the gray region in 
Fig.~\ref{fig:Powers_UK}~a), and is recovered at larger values of $\delta$, where the 
stable operating state strongly differs from the unperturbed one. 
Synchrony is lost for $\delta \nearrow 2.1$ and $\delta \searrow 5.2$ as the power flow solutions become unstable
($\lambda_2\rightarrow0$), similarly to results reported in Ref.~\cite{Wittahaut-EPJS}.

The added line labeled by $(b)$ [see Fig.~\ref{fig:UK}] is chosen to connect two nodes such that 
$\theta_{N,S}\approx-0.9\pi\in[-\pi,-\pi/2]$. As can be seen in Fig.~\ref{fig:Powers_UK}~b) power is flowing from 
the south to the north along this new line. Despite the associated reduction of the power flow on
the two original lines, the Lyapunov exponent increases
as predicted in Fig.~\ref{fig:Summary}. For larger added capacity, however,
$\lambda_2$ reaches a maximum, then starts to decrease, and synchrony is never lost.
This observation can be understood qualitatively in terms of our simple model:
as the capacity of the new connection increases, the difference $\theta_{N,S}$, originally in 
the $3^\textrm{rd}$ quadrant increases until eventually it reaches the $4^\textrm{th}$ trigonometric quadrant 
for which the correction to $\lambda_2$ is expected to become negative.

We finally add line $(c)$ [see Fig.~\ref{fig:UK}] between
two nodes with $\theta_{N,S}\approx0.3\pi$. This time
power flows through the new line from the north to the south. The loads on the original connections between 
the two zones, which were already close to saturation, increase further.
Quite interestingly, the linear stability of the solution is improved, $\Delta \lambda_2 \le 0$,
despite this load increase. 
The solution followed as $\delta$ is raised from 0 remains linearly stable in the whole capacity range investigated
$\delta\in[0,13]$. When the capacity of the additional line reaches $5.98$ units of power, the numerical simulations
also converge to another stable solution of the power flow equations [see Fig.~\ref{fig:Powers_UK}~c)].
The behavior of $\lambda_2$ indicates that the new, large$-\delta$ solution becomes unstable ($\lambda_2=0$) 
when $\delta \searrow 5.98$.
The regime $\delta\geq 5.98$ is an example of coexistence of multiple stable power flow solutions 
\cite{Delabays2015,Korsak, Janssens, Ochab, Mehta}.


\begin{figure}[htbp]
 \centering
 \includegraphics[width=\columnwidth]{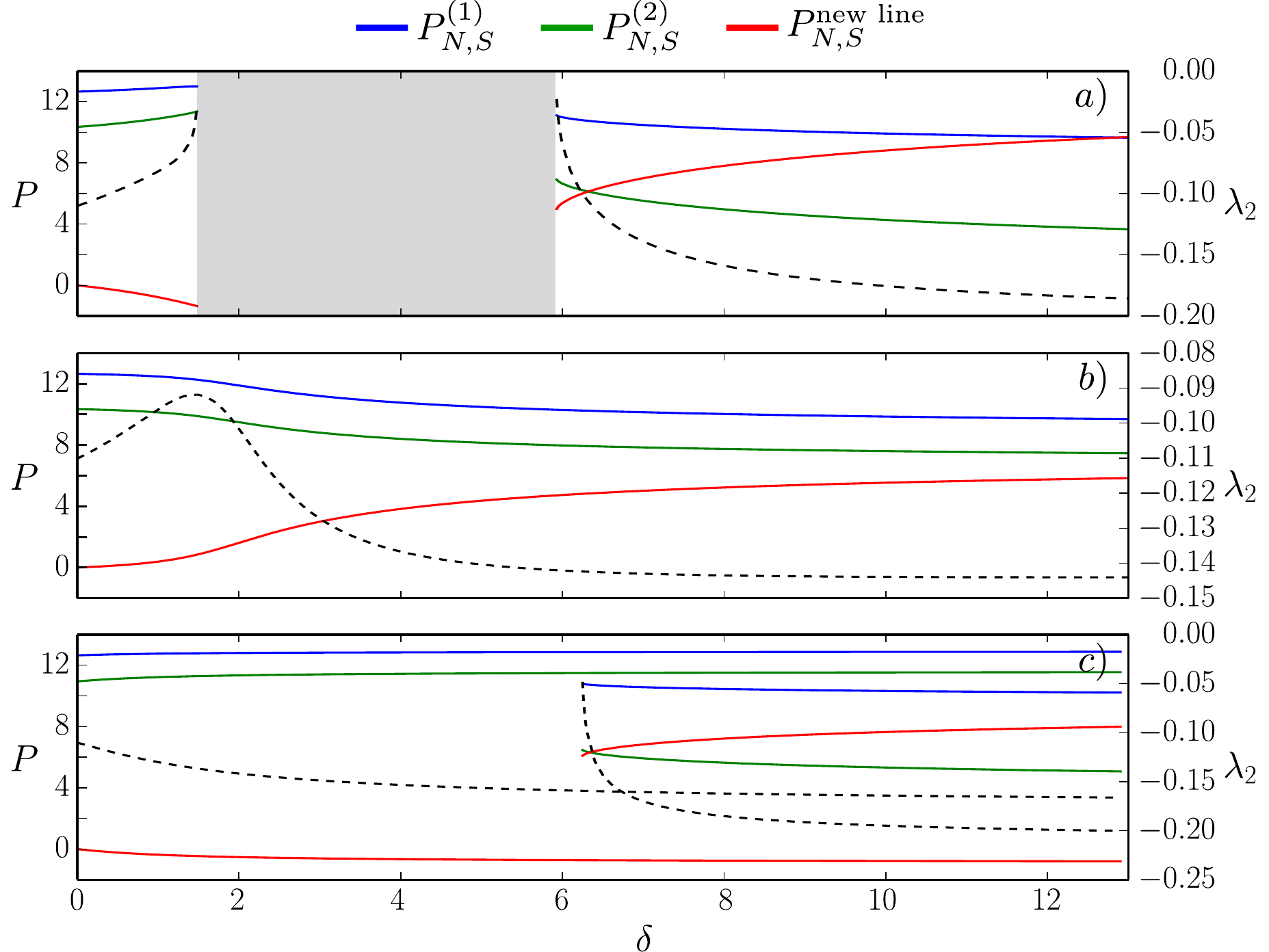}
 \caption{(Color online)  Lyapunov exponent (dashed) and power flowing though the lines 
 connecting the north and south areas as a function of the capacity of the additional line $\delta$,
 for the UK transmission grid in the case of line 
 capacities uniformly distributed in the interval $[9.75,16.25]$.
 Each panel refers to one of the line additions represented in Fig.~\ref{fig:UK} 
 and illustrates one of the three Braess scenarios identified in this work.}
 \label{fig:Power_UK_Kvariable}
\end{figure}

To assess the robustness of the perturbation theory results we repeat the numerical simulations for the UK 
transmission grid including line capacity variations of $\pm25\%$ with respect to the uniform case, $K\equiv13$,
presented above.
We take line capacities uniformly distributed in the interval $[9.75,16.25]$ keeping the most loaded 
line crossing the north-south border at $K=13$, and consider the same network upgrades discussed earlier. 
The numerical results presented in Fig.~\ref{fig:Power_UK_Kvariable} are very similar to those of 
Fig.~\ref{fig:Powers_UK}, indicating that the three different Braess scenarios identified for the uniform 
line capacity case are robust with respect to the significant capacity variations considered.

\section{Conclusion.}

We classified the impact of a line addition in an AC power grid into four 
possible scenarios depending on the change in linear stability of the synchronous solution and 
on the change in power load on the lines. 
For the chain model, we showed that the effect of such network upgrades 
depends uniquely on the voltage angle difference between the nodes connected by the
new line. This classification is summarized in Fig.~\ref{fig:Summary}, and we
showed that it can be extended to meshed networks having the topology of transmission grids.
In this case it is however less straightforward to predict from the unperturbed operational state which of the
four scenarios will be realized.

We think that our theory has significantly deepened our understanding of the Braess paradox in electric
power systems. More generally, it is based on rather generic models, which 
suggests that Braess paradoxes, in the form of weaker stability of the synchronous state after 
coupling addition, are ubiquitous in systems of coupled oscillators. 
Future works should attempt to extend this theory to the non pertubative regime of large line capacity
and include dissipation effects which become important as voltage angle differences become large.

We thank R.~Delabays for useful discussions.
This work was supported by the Swiss National Science Foundation.

\appendix
\section{Jacobi matrices}\label{sec:Jacobi matrices}
\subsection{Properties of Jacobi matrices.}
In this section we list some of the properties of the eigenvalues and eigenvectors of Jacobi matrices.
For further details and proofs of these results see Ref.~[\onlinecite{Jacobi_matrices}].
Consider the following positive semi-definite, symmetric, tridiagonal matrix with strictly negative subdiagonals
\begin{equation}
  J=\left(
 \begin{array}{cccccc}
  a_1  &  -b_1  &   0    & \ldots &     0 \\
 -b_1  &   a_2  & -b_2   &   0    &   \vdots \\
   0   & \ddots & \ddots & \ddots &   0 \\
\vdots &        & \ddots & \ddots &  -b_{n-1} \\
   0   & \ldots &   0    & -b_{n-1} & a_n \\
 \end{array}
 \right)\,, \qquad b_i>0 \,.
\end{equation}
Such matrices are also known as Jacobi matrices and have the property that 
their eigenvalues are distinct [\onlinecite{Jacobi_matrices}].
Furthermore, the principal minors of $J$ (the $k^\textrm{th}$ principal minor $J$ is the truncated version of
$J$ consisting of $J_{i,j}$ with $i,j=1,\ldots,k\leq n$) satisfy the following recurrence relation
\begin{equation}\label{eq:recurrence relation}
 D_{k+1}(\lambda)=(a_{k+1}-\lambda)D_{k}(\lambda)-b_k^2D_{k-1}(\lambda)\,,
\end{equation}
where $D_{k}(\lambda)$ is the characteristic polynomial of the $k^\textrm{th}$ principal minor of $J$ ($D_0(\lambda)=1$ and $D_1(\lambda)=a_1-\lambda$).
In particular $D_{n}(\lambda)$ is the characteristic polynomial of $J$ which vanishes when
$\lambda$ is equal to one of the eigenvalues $\lambda_1<\lambda_2<\ldots<\lambda_n$ of $J$.
It can be shown [\onlinecite{Jacobi_matrices}] that the sequence
\begin{equation}
 \left\{D_{n-1}(\lambda),D_{n-2}(\lambda),\ldots,D_{1}(\lambda),D_{0}(\lambda)\right\}
\end{equation}
contains $j-1$ sign changes when evaluated at the $j^\textrm{th}$ eigenvalue $\lambda=\lambda_j$.

If $\mathbf{u}^{(j)}=(u^{(j)}_1,\ldots,u^{(j)}_n)$ is the eigenvector of $J$ associated to $\lambda_j$
it is straightforward to show that the coefficients of $\mathbf u^{(j)}$ satisfy a recurrence relation 
which is similar to that of Eq.~(\ref{eq:recurrence relation})
\begin{equation}\label{eq:recurrence relation eigenvectors}
\begin{array}{c}
 J\mathbf{u}^{(j)}=\lambda_j\mathbf{u}^{(j)} \,,\\[2mm]
\Leftrightarrow -b_{k-1}u^{(j)}_{k-1}+a_{k}u^{(j)}_k-b_ku^{(j)}_{k+1}=\lambda_ju^{(j)}_k \,, \\[2mm] 
\Leftrightarrow b_ku^{(j)}_{k+1}=(a_{k}-\lambda_j)u^{(j)}_k-b_{k-1}u^{(j)}_{k-1} \,, 
\end{array} 
\end{equation}
for $k\in\{1,\ldots,n\}$ and $u^{(j)}_{0}=u^{(j)}_{n+1}=0$. In fact one obtains that (\ref{eq:recurrence relation eigenvectors})
is fulfilled by
\begin{equation}
 u^{(j)}_k\propto \frac{D_{k-1}(\lambda_j)}{b_1\ldots b_{k-1}} \,.
\end{equation}
Hence, given the sign property of the sequence $\{D(\lambda_j)\}$, it is clear that the components of 
the eigenvector $\mathbf{u}^{(j)}$ will have $j-1$ sign changes.

We mention a last property [\onlinecite{Jacobi_matrices}] of the eigenvectors of $J$ which is useful for our electrical model. Given two eigenvectors $\mathbf{u}^{(j)}$
and $\mathbf{u}^{(i)}$ we have
\begin{equation}
 \begin{array}{c}
 -b_{k-1}u^{(j)}_{k-1}+a_{k}u^{(j)}_k-b_ku^{(j)}_{k+1}=\lambda_ju^{(j)}_k \,,\\[2mm]
 -b_{k-1}u^{(i)}_{k-1}+a_{k}u^{(i)}_k-b_ku^{(i)}_{k+1}=\lambda_iu^{(i)}_k\,.
 \end{array}
\end{equation}
Eliminating $a_k$ one obtains
\begin{equation}
 b_k(u^{(i)}_ku^{(j)}_{k+1}-u^{(i)}_{k+1}u^{(j)}_{k})+b_{k-1}(u^{(i)}_{k}u^{(j)}_{k-1}-u^{(i)}_{k-1}u^{(j)}_{k})=(\lambda_i-\lambda_j)u^{(j)}_k u^{(i)}_k\,,
\end{equation}
which summed over $k=1,2,\ldots,l$ gives
\begin{equation}\label{eq:relation between two eigenvectors}
 b_l(u^{(i)}_lu^{(j)}_{l+1}-u^{(i)}_{l+1}u^{(j)}_{l})=(\lambda_i-\lambda_j)\sum_{k=1}^l u^{(j)}_k u^{(i)}_k\,.
\end{equation}
\subsection{Connection to the chain model.}
The matrix $-M$ constructed for our chain model prior to the line addition
is a Jacobi matrix with the additional property that since $a_i=b_i+b_{i-1}$ its first eigenvalue $\lambda_1$ is equal to zero and 
the corresponding eigenvector is $\mathbf{u}^{(1)}=(1,\ldots,1)$.
Thus, applying Eq.~(\ref{eq:relation between two eigenvectors}) to $-M$ with $j=1$ and $i=2$ yields
a relation between any two consecutive components of $\mathbf{u}^{(2)}$
\begin{equation}\label{eq:relation between components of u2}
 (u^{(2)}_l-u^{(2)}_{l+1})=\frac{\lambda_2}{b_l}\sum_{k=1}^l u^{(2)}_k\,.
\end{equation}
Additionally, according to the properties of Jacobi matrices, there will be only one sign change
in the list of the coefficients of $\mathbf{u}^{(2)}$ which therefore will have the form $(\pm,\ldots,\pm,\mp,\ldots,\mp)$. 
Lastly, the orthogonality relation between $\mathbf{u}^{(1)}$ and $\mathbf{u}^{(2)}$ implies that
\begin{equation}\label{eq:Orthogonality condition}
 \mathbf{u}^{(1)\top}\mathbf{u}^{(2)}=0 \quad \Leftrightarrow \quad \sum_{k=1}^n u^{(2)}_k = 0\,.
 \end{equation}
Using Eq.~(\ref{eq:Orthogonality condition}) 
and the sign properties of the coefficients of $\mathbf{u}^{(2)}$ suffices to see that Eq.~(\ref{eq:relation between components of u2})
leads to the conclusion that the components of $\mathbf{u}^{(2)}$ are monotonously ordered (i.e. for any $i> j$ we either have $u_i^{(2)}\geq u_j^{(2)}$ or $u_i^{(2)}\leq u_j^{(2)}$).

\section{Perturbation theory for a generic graph}\label{sec:Perturbation generic graph}
In this section we extend the calculation of the leading order correction to the Lyapunov exponent $\lambda_2$ 
resulting from the addition of a new line to the case of a generic electric network.
Given a generic network of $N$ nodes and lines of capacity $K_{i,j}$, 
let $\left\{\theta_i\right\}$ and $\{\tilde{\theta}_i\}$ respectively denote 
the solutions of the power flow equations~(\ref{eq:PowerFlow})
before and after the addition of a line of capacity $\delta\ll K_{i,j}$ between nodes $\alpha$ and $\beta$.
Assuming the $\tilde{\theta}_i$'s are small deviations of the unperturbed solution
($\tilde{\theta_i}\approx\theta_i+\delta\theta_i$ with $|\delta\theta_i|\ll 1$),
we expand the power flow equations to leading order in $\delta$
\begin{equation}\label{eq:Expansion power flow}
 \begin{array}{l}
 \displaystyle 0=\sum_ {l \sim i}K_{l,i}\cos(\theta_{l}-\theta_{i})(\delta\theta_l-\delta\theta_i) \quad i\neq\alpha,\beta\,, \\
 \displaystyle 0=\sum_{l\sim \alpha,\,l\neq\beta}K_{l,\alpha}\cos(\theta_{l}-\theta_{\alpha})(\delta\theta_l-\delta\theta_\alpha) +\delta\sin(\theta_\beta-\theta_\alpha)\quad i=\alpha\,, \\
 \displaystyle 0=\sum_{l\sim \beta,\,l\neq\alpha}K_{l,\beta}\cos(\theta_{l}-\theta_{\beta})(\delta\theta_l-\delta\theta_\beta) +\delta\sin(\theta_\alpha-\theta_\beta) \quad i=\beta\,. \\
 \end{array}
\end{equation}
Eqs.~(\ref{eq:Expansion power flow}) can be rewritten using the stability matrix, 
$M$ [defined below Eq.~(\ref{eq:Eigenvalue Eq})], 
of the system prior to the line addition
\begin{equation}\label{eq:delta theta}
 M \pmb{\delta\theta}=\delta\sin{\theta_{\alpha,\beta}} \pmb{v}\,,
\end{equation}
where $\pmb{\delta\theta}=(\delta\theta_1,\hdots,\delta\theta_N)$ and $\pmb{v}$ is the $N$ dimensional vector whose $i^\textrm{th}$ 
component is equal to $v_i=\delta_{i,\alpha}-\delta_{i,\beta}$.

$M$, being a real symmetric matrix, is diagonalized by an orthogonal matrix $T$ whose $l^\textrm{th}$ column is the 
 $l^\textrm{th}$ eigenvector, $\mathbf{u}^{(l)}$, of $M$. 
Furthermore, the $U(1)$ symmetry of the power flow equations implies that one of the eigenvalues
of $M$ is null ($\lambda_1=0$ assocuated to ${\bf u}^{(1)}=(1,\ldots,1)/\sqrt{N}$).
Since $M$ is singular it cannot be inverted. However, Eq.~(\ref{eq:delta theta}) can be solved for the $\delta\theta_i$'s 
by using the Moore-Penrose pseudoinverse of $M$ defined as
\begin{equation}\label{eq:Definition pseudo inverse}
  M^{-1} =T\left(
 \begin{array}{cccc}
 0& & & \\
  &\lambda_2^{-1}&& \\
  & &\ddots& \\
  & & &\lambda_N^{-1}
 \end{array}\right) T^\top \,,
\end{equation}
where $T=(\mathbf{u}^{(1)},\ldots,\mathbf{u}^{(N)})$ and the $\lambda_l$'s are the eigenvalues of $M$.
$M^{-1}$ is such that $M^{-1}M=M^{-1}M=\mathbb{1}-\mathbf{u}^{(1)}{\mathbf{u}^{(1)}}^\top $, where $\mathbf{u}^{(1)}{\mathbf{u}^{(1)}}^\top$ is equal to 
the $N\times N$ matrix having $1/N$ for all it's entries.
Multiplying (\ref{eq:delta theta}) by $M^{-1}$ yields
\begin{equation}
 \left(\begin{array}{c}
  \delta\theta_1\\
  \vdots \\
  \delta\theta_N
 \end{array}\right)
 -\frac{1}{N}
  \left(\begin{array}{c}
  \sum_l\delta\theta_l\\
  \vdots \\
  \sum_l\delta\theta_l\\
 \end{array}\right)=
 \delta\sin{\theta_{\alpha,\beta}}
  \left(\begin{array}{c}
  M^{-1}_{1,\alpha}-M^{-1}_{1,\beta}\\
  \vdots \\
  M^{-1}_{N,\alpha}-M^{-1}_{N,\beta}
  \end{array}\right)\,.
\end{equation}
The difference of $\delta\theta_i$'s between any two nodes is given by
\begin{equation}\label{eq:delta theta 2}
 \delta\theta_i-\delta\theta_j\equiv\epsilon_{i,j}=\delta\sin\theta_{\alpha,\beta}\left[\left(M^{-1}_{i,\alpha}-M^{-1}_{i,\beta}\right)-\left(M^{-1}_{j,\alpha}-M^{-1}_{j,\beta}\right)\right]\,,
\end{equation}
where the term $\sum_l\delta\theta_l$ drops due to the global rotational invariance of the power flow solution.
Finally, Eq.~(\ref{eq:delta theta 2}) can be expressed in terms of the eigenvectors of $M$ making use of (\ref{eq:Definition pseudo inverse}).
This yields
\begin{equation}\label{eq:delta theta 3}
 \epsilon_{i,j}=\delta\sin\theta_{\alpha,\beta}\sum_{l\geq2}\left(u^{(l)}_i-u^{(l)}_j\right)\left(u^{(l)}_\alpha-u^{(l)}_\beta\right)\lambda_l^{-1}\,,
\end{equation}
for any node $i$ connected to node $j$.
Having established the correction to the power flow solution (\ref{eq:delta theta 3}), it is straight forward to compute 
the leading correction to the stability matrix, $\Delta M$, and to obtain the correction of the Lyapunov exponent 
$\Delta\lambda_2={\mathbf{u}^{(2)}}^\top\Delta M\mathbf{u}^{(2)}$. 
The final expression for $\Delta\lambda_2$ in the case of a generic network is presented in 
Eq.~(\ref{eq:DeltaLambda2 generic}).

\bibliography{bibliography}
\end{document}